\documentclass{aa}
\usepackage{amssymb}
\usepackage{times}
\usepackage{graphicx}
\usepackage{hyperref}

\def\asec{\ifmmode ^{\prime\prime}\else$^{\prime\prime}$\fi}

\def\it{\sl}
\def\degs{\ifmmode ^{\circ}\else$^{\circ}$\fi}
\def\amin{\ifmmode ^{\prime}\else$^{\prime}$\fi}
\def\asec{\ifmmode ^{\prime\prime}\else$^{\prime\prime}$\fi}
\def\fhour{\hbox{$.\!\!^{\rm h}$}}           % Fractions of h

\def\wga{1\,WGA\,J1958.2+3232\,}

\begin{document}

\title{ On the  Orbital Period of the  Intermediate Polar \\  1WGA J1958.2+3232} 

\author{Sergei V. Zharikov$^1$
\and  Gaghik H. Tovmassian$^1$
\and Juan Echevarr\'{\i}a$^2$}

\institute{ $^1$  Observatorio  Astron\'{o}mico  Nacional,   Instituto  de
Astronom\'{\i}a, UNAM,  22800, Ensenada, B.C., M\'{e}xico
\thanks{use
for  smail  P.O.  Box  439027,  San  Diego,  CA,  92143-9027,  USA}  \\
$^2$ Instituto de Astronom\'{\i}a, UNAM, Apartado Postal 70-264, 
04510 M\'{e}xico, D.F., M\'{e}xico}
\authorrunning{S.  V.  Zharikov et  al.,  } \titlerunning{The  Orbital
Period     of     1WGA     J1958.2+3232}    \offprints{Zharikov     S}
\mail{zhar@astrosen.unam.mx}  \date{Received   --  /  Accepted   --  }

\abstract{
Recently,   Norton   et  al.   (\cite{Norton}),    on   the  basis   of
multiwavelength photometry of \wga\,, argued that the\ --1 day alias of
the strongest peak in the power spectrum is the true orbital period of
the system, casting doubts on the period estimated by Zharikov et al.
(\cite{Zharikov}).  We re-analyzed this system using our
 photometric and spectroscopic
data  along with  the data kindly  provided by Andy
Norton  and confirm our  previous finding.  After refining our analysis we
 find that the true
orbital period of this binary system is 4\fhour35.
}
\maketitle

\section{Introduction.}

Israel et  al.  (\cite{Israel1}) discovered that \wga\,  was a pulsating
X-ray source. 
Strong modulations of  this source in X-rays were  obtained from the ROSAT
PSPC ($721\pm14$ sec)  and a more accurate period of $734\pm1$ sec
from ASCA was
presented  by  Israel  et  al.   (\cite{Israel1}) and  Israel  et  al.
(\cite{Israel2}).  Photometric observations of the optical counterpart
of \object{\wga} exhibited  strong optical variations, compatible with
the X-ray  (within 12 min) period (Uslenghi  et al.  \cite{Uslenghi}).
This modulation was interpreted as an evidence of the spin period of the WD
in a close binary system.  Uslenghi et al. (\cite{Uslenghi}) detected a
circular  polarization from  the source  in the  R and  I  bands, with
 evidence  for a possible  modulation of  the polarization  at  twice the
previously    observed   pulsation    period.  
\wga was announced as  a new Intermediated  Polar (IP) by
Negueruela  et  al.  (\cite{Negueruela}) from  spectral  observations.
  Zharikov    et   al.
(\cite{Zharikov})  obtained   time  resolved spectroscopy and 
R-band   photometry from  which they deduced  an orbital period  of 4\fhour36
and confirmed the pulsation period of 733\,sec.  Later on, Norton et al.,
(\cite{Norton})  obtained  UBVRI photometry  and  reported  that the  orbital
period  was $5.387\pm0.006$ hours, corresponding to the\  --1  day alias  
of the  period found by Zharikov  et  al.  (\cite{Zharikov}).    
They  had  some  ambiguity  in
determining  which  of  the  daily  cycle  aliases  of  low (orbital) frequency
  and  intermediate (beat) frequency   to  pick  up,  because
selecting the strongest  peak in low frequencies was  forcing the beat
period  into  a\ --2  day  alias  of  the intermediate  frequency  peak.
Through detection  of the beat  frequency, Norton et  al. (\cite{Norton})
also confirmed that the rotational  period of the white dwarf is twice
the  pulse  period,  and  they  confirmed  the  presence  of the circular
polarization in the source by detecting oppositely signed polarization
in each of the B and R bands.

In  this  letter,  we  re-analyze  our spectral  and  photometric  data
together with photometric data from Norton et al. (\cite{Norton})
confirm and refine our previous period estimate of 4\fhour35.

\section{Combined data and search of period}

The UBVRI data of  the optical counterpart of 1WGA\,J1958.2+3232
were obtained by  Norton et al. (\cite{Norton})  on 9-15 July
2000.  The R-band  time-resolved photometry of
Zharikov et al. (\cite{Zharikov}) was  obtained on August of 2 and
3. We also obtained   time-resolved spectroscopy of \wga on
4-6  Aug.   2000.   Details  of the    observations  are   provided  in
corresponding papers. It is important  to note that the total duration
of our  spectroscopic  observations on  the second night was  7\fhour7, thus
covering almost  two  orbital  periods.  A total of  68 spectra
were obtained (Zharikov et al. \cite{Zharikov}).

\begin{figure}[h]
\includegraphics[width=8.5cm,clip=]{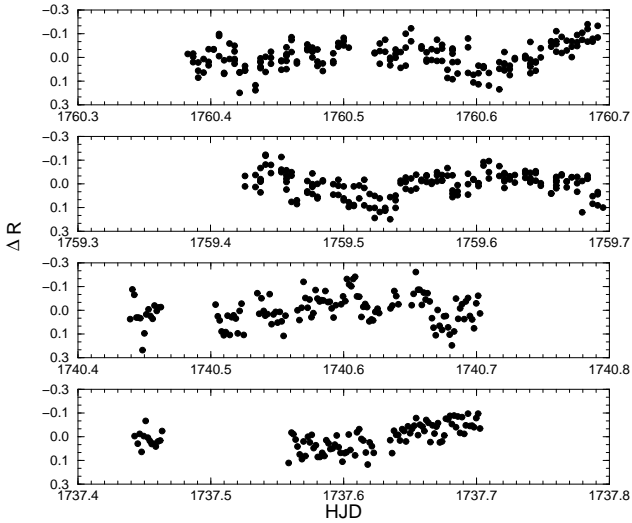}
\caption{\wga\  light curves in the $R_{\mathrm c}$ band  are presented. 
The $\mathrm{HJD = 2541000+  1737}$ and $1740$ corresponds to  
Norton et al. (\cite{Norton}) (low panels);
the other 2 nights (upper panels) are  from Zharikov et  al. (\cite{Zharikov}).
 }
\label{fig1}
\end{figure}

 As a  first step to a verify  the binary system orbital period,  we 
combined the R-band data from both  data sets.  The light curves of \wga\,
in the $R_c$ band are presented in Figure.\ref{fig1}.  From this figure we
can  see  similar  behavior of both lightcurves.  
However, our  time coverage  is somewhat longer and data spacing is more 
even and more dense.

\begin{figure}[h]
\includegraphics[width=8.5cm,clip=]{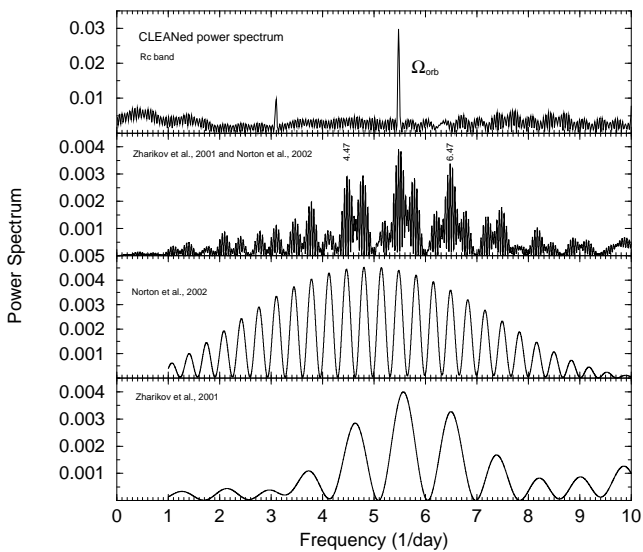}
\caption{The  low frequency end of the  power  spectra of  $R_{\mathrm c}$ light 
curves are
given. The top panel is a CLEANed power spectrum of the $R_{\mathrm c}$ light 
curve from combined 
Zharikov  et al.  (\cite{Zharikov})  and Norton  et al.  (\cite{Norton})
data.}
\label{fig2}
\end{figure}

The  photometric  data  were  analyzed  for  periodicities  using  the
Discrete Fourier Transform code (Deeming \cite{Deeming}) with a CLEAN
procedure (Roberts  et al. \cite{Roberts}). The power  spectrum at low
frequencies is presented in  Figure.\ref{fig2}. The power spectrum of our
$R_{\mathrm c}$ data  and Norton et al.  (\cite{Norton}) $R$ data  are given 
separately in the lower panels.   
The power spectrum of combined data are  presented in the second  from the
top panel. The largest peak  $\Omega=5.47\ \mathrm d^{-1}$ and its $\pm 1$ day aliases
are marked.  The top panel is a CLEANed power spectrum of the combined
$R_{\mathrm c}$   data.  The   CLEANed  power   spectrum  shows a  peak  at
$\Omega=5.4734054\pm    0.0215067\ \mathrm d^{-1}$,    corresponding   to
$P=0.1827016\pm 0.000715\ \mathrm d $. We  note here that CLEAN will always
clean data to the highest peak in the power spectrum, so on its own this
is not a true test of which of the 1-day aliases is the correct one, but 
CLEAN helps to determine  the highest  frequency exactly. 

\begin{figure}[h]
\includegraphics[width=8.cm,clip=]{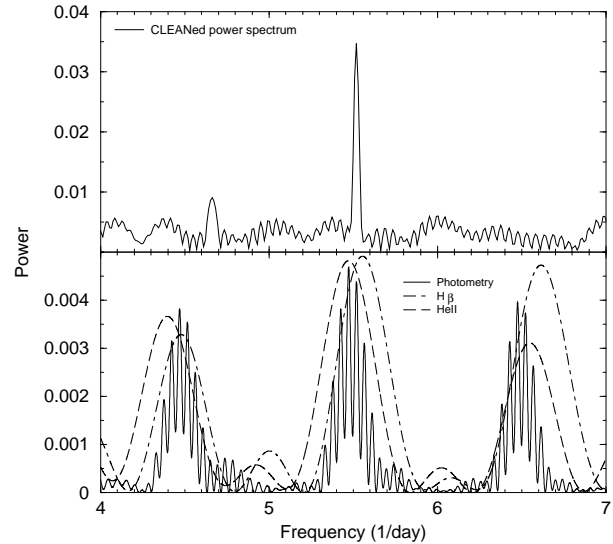}
\caption{The CLEANed  power spectrum of AFD photometrical data
from Zharikov et al.  (\cite{Zharikov}) and Norton et al. (\cite{Norton})
(top  panel). 
On the lower panel the uncleaned power spectrum of the AFD is shown.
 The aliases ($\pm 1\ \mathrm d^{-1}$) are presented too.
The power spectra of the RV variations of and \ion{He}{ii} 4686
and \ion{H$_\beta$}{}  are overplotted. They are scaled
to the  amplitude of  the power spectrum  of photometry.   The maximum frequency peak  corresponds  to the orbital  period of  the system.
}
\label{fig3}
\end{figure}

After  this,  we tested the photometric  data  including all  other
filters.  We subtracted the average magnitude from the photometric data of
each night of observations and merged  all data in one set.  The power
spectrum   resulting from the all-filter photometric data  (AFD) is presented
in  Figure.\ref{fig3} (lower  panel).  The  maximum peak  corresponds to a
$\sim 5.52 \mathrm{ (d^{-1})}$ frequency.  Naturally one day  aliases also come
up with lower amplitudes.

We again applied the  CLEAN procedure which  is aimed to distinguish  the 
alias periods originating from uneven  distribution of data and works nicely
on large data  sets containing well defined alias  periods. The
power  spectrum of  the AFD set (top  panel in  Figure.\ref{fig3})  again
shows   a  single peak  at $\Omega_{\mathrm o}  =  5.518908\pm 0.010315\
\mathrm d^{-1}$,   which corresponds  to   $P =  0.181195\pm 0.000339\
\mathrm d$ (4\fhour35 period).

\begin{table*}[t]
\caption{The parameters of the $\sin$ fit of RV data.}
\begin{tabular}{|c|cccc|cccc|} \hline
Line  &\multicolumn{4}{|c|}{ \ion{He}{ii}} & \multicolumn{4}{c|}{\ion{H$_{\beta}$}} \\ \hline
$\Omega$  &$\Omega_o$   &$\Omega_o-1$  &$\Omega_o+1$  &$\Omega_N$  &$\Omega_o$ &$\Omega_o-1$  &$\Omega_o+1$  &$\Omega_N$    \\ \cline{2-9}
 ($\mathrm d^{-1}$)& 5.5189    &  4.5189      & 6.5189    & 4.455    & 5.5189    &  4.45189   & 6.5189  & 4.455 \\ \hline
P (d)           & 0.18120    &   0.22129    & 0.15340       &0.22447    & 0.18120          & 0.22129    & 0.15340 &0.22447  \\\hline\hline 
$\gamma_0^*$(km/s)& -72.1750      &   -77.8      & -97.2         &-66.63     &-38.07      &-45.39    &-31.92     &-42.2588 \\
$K_1^{**}$ (km/s)      & -189.03       &   -167.18    & 169.7         &-176.73    & -74.28     & 70.77    &56.38      &-71.0629 \\
$t_0^{***}$ (HJD)   &  51763.5537   &   51764.1556 & 51765.9766    &51764.1750 & 51763.8587 &51763.5302& 51763.6224& 51763.8733\\ \hline
$\chi^2$        & {\bf 140.3/57}& 401.8/57     & 374.1/57      &322.7/57   & {\bf 80.2/70} & 94.23/70 &125.35/70& 86.4/70 \\ 
$\sigma$        & { \bf 68.62 }  & 113.68       & 95.66         &100.14    & {\bf 43.37}   &   47.57  & 53.97    &45.40 \\ \hline  
\end{tabular}\\  
$^* \gamma_o$ is the
systematic velocity of the system \\
$^{**}K_1$ is the semi-amplitude of
the radial  velocity \\
$^{***} 2540000+t_o$
\label{tab}
\end{table*}

However, the  crucial and the  most unambiguous  confirmation of  the 4\fhour35
orbital period  comes from the  consideration of radial  velocity (RV)
data previously obtained  by us. The methods used to measure the radial
velocities in \ion{H$_\beta$}{} and  \ion{He}{ii}  were described by Zharikov et al.
(\cite{Zharikov}).  The power spectra of RV data from Zharikov
et  al. (\cite{Zharikov})  are overplotted  in Figure  \ref{fig3}.  They
show wide  peaks coinciding with  the photometric results.
 While the spectroscopic data
 do not allow a  precise determination of the orbital period, they
were derived  from three consecutive nights  of prolonged observations
covering more  than one  orbital period, which  allows us to  test the
$\pm 1$ day period aliases in the power spectra on the actual data.

\begin{figure*}[t]
\includegraphics[width=8.5cm,clip=]{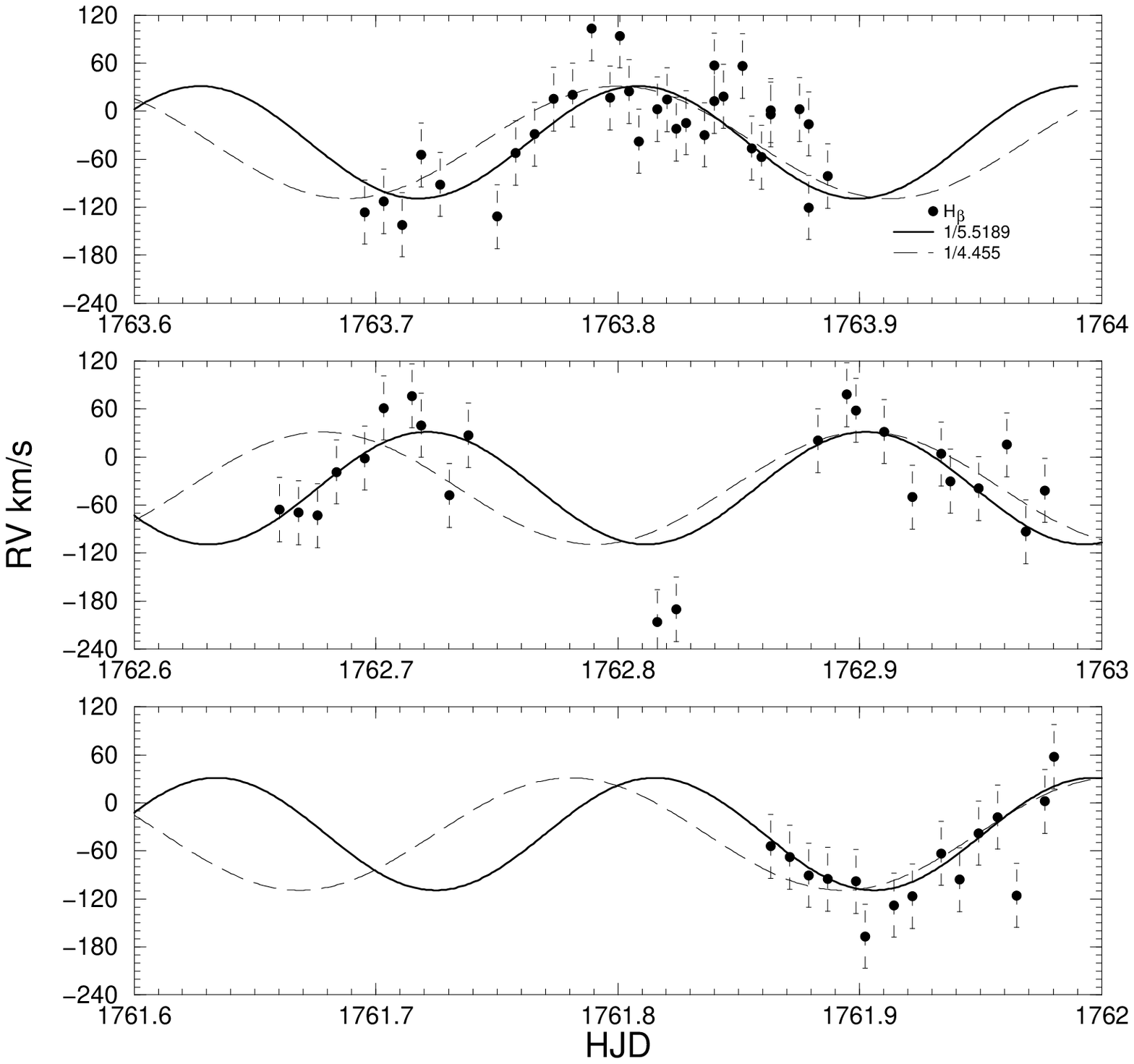}
\includegraphics[width=8.5cm,clip=]{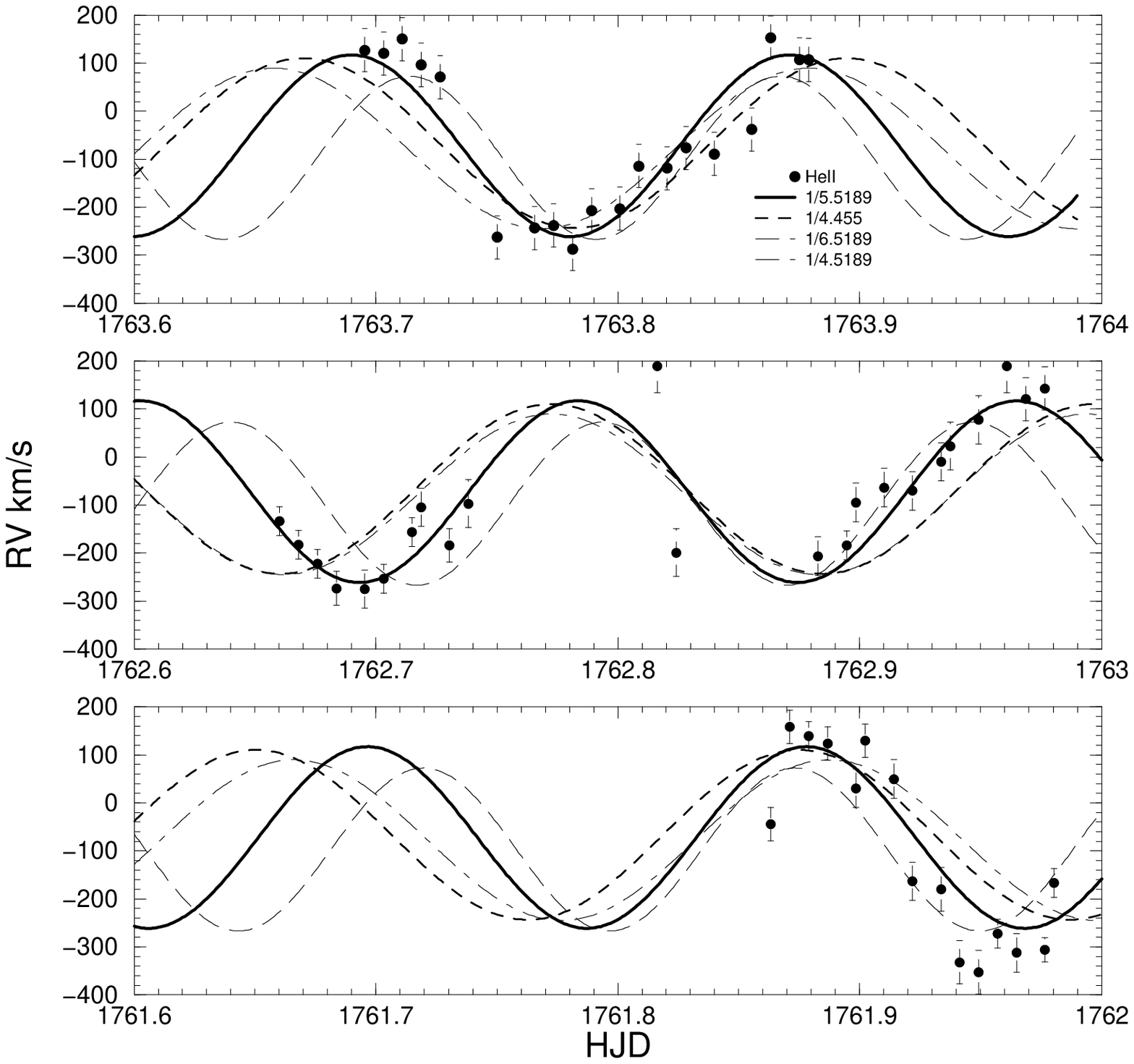}
\caption{The  radial velocity measurements  of the  
emission lines  of \ion{He}{ii}
4686 and \ion{H$_\beta$}{}  for each night of observations.
 The curves
correspond to   $sine$ fits to the radial velocity data  with 
estimated  orbital period  and its $\pm 1\ \mathrm
d^{-1}$ aliases. The solid line is the best fit with the $4\fhour35$ period.}
\label{fig4}
\end{figure*}

 In Figure \ref{fig4} we present unfolded radial velocity measurements
of the  emission lines of  \ion{He}{ii} 4686 and  \ion{H$_\beta$}{} at
each  night  of  observations.   The  errors of  RV  measurements  are
presented in  corresponding panels. Fits of a $sine$ function   to the data
 with the period estimated by us are overplotted as a
solid line. The $\pm 1$ aliases are shown as a thin dashed lines.  The\ \
--1 day  (1/4.455)    alias   selected    by    Norton   et
al.  (\cite{Norton}) as  a true  orbital period  and drawn  with thick
dashed  line can  not give  a satisfactory  fit to  the data  from the
second  night,  where  almost  two  orbital periods  were  covered  by
the observations.
The results of  the  $\chi^2$  fit by
$$ v(t)  = \gamma_o+K_1\sin(2\pi(t-t_o)/P),$$ where 
$\gamma_{\mathrm o}$, $K_1$ and  $t_{\mathrm o}$ were free parameters 
 for  our best orbital period  estimate $\Omega_{\mathrm o}$,
its $\pm  1$ day  aliases and orbital  period $\Omega_{\mathrm N}$ by  Norton et
al. (\cite{Norton}) are  given at  Table \ref{tab}.   The best  fit  result was
obtained for \ion{He}{II} RV data at frequency $\Omega_{\mathrm o} = 5.5189\ \mathrm d^{-1}$ 
significantly  exceeding fits  with other  frequencies.  The  results for
\ion{H$_\beta$}{} are less conclusive due to the smaller amplitude and
larger errors of  the RV measurements.  However, in  this case also we
can see that  at $\Omega_{\mathrm o}$ we have the lowest  values of  $\chi^2$
and $\sigma$.

Not surprisingly, the $\chi^2$  and $\sigma$ values from Table 1 confirm what
can be seen  with the naked eye, that the  period corresponding to the
strongest peak in the power spectrum is most probably the true orbital
period of the  system.  We adopted $P_{\mathrm{orb}} =  0.181195\pm 0.000339\ \mathrm d$ as
the final value  for the orbital period of \wga. A longer time base of
spectroscopic observations is needed to improve this value.

\section{Conclusion}

Norton  et  al. (\cite{Norton})  chose  the $\Omega_{\mathrm N}  =  4.455\pm0.005$\
$\mathrm d^{-1}$,  or $P_{\mathrm N}=5.387\pm  0.006\ \mathrm h$,  as  the orbital  period of  the
system from the analysis of the  power spectrum  peak strength combination.
They noted  that the power  spectrum is dominated by  three sets of
signals at $\sim 5.5\ \mathrm d^{-1}$,  $55.5\ \mathrm d^{-1}$ and $117.8\ \mathrm d^{-1}$ but
the strongest peaks in each of the three sets are not harmonically related
to  each  other.   The   solution  $\Omega_{\mathrm N}$  was  selected  as the  more
probable. They assume that more extreme aliases combinations
are  unlikely,  since  the  power at  these  alias are  low,
although  such combination  are not excluded.   
In our opinion the strength of peaks of power spectra are highly
dependent on the quality of the data and sampling. The photometric
data of Norton et al. (\cite{Norton})  is certainly undersampled for such
far-reaching conclusions. On the other hand, the spectroscopic observations
 presented here unambiguously identify the orbital period
of the system.

\begin{figure}[t]
\includegraphics[width=8.5cm,clip=]{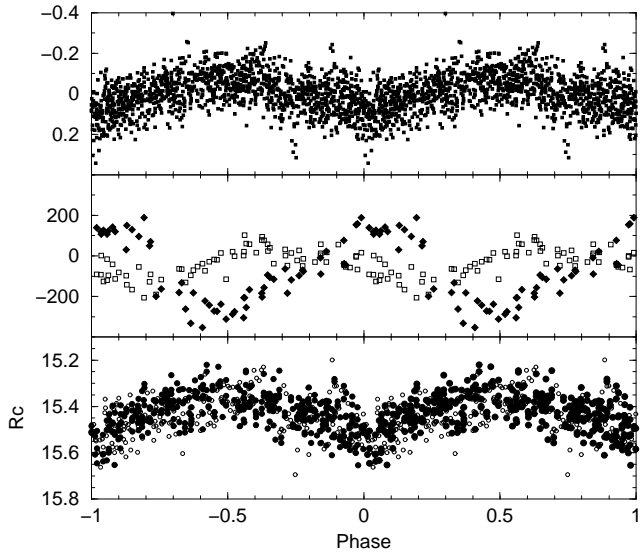}
\caption{The  radial  velocity  curves   of
\ion{H$_\beta$}{}  and \ion{He}{ii} 4686,  folded with  the spectroscopic
orbital   period  of $4\fhour35$,   are   presented in the  middle   panel.  
The combined   $R_\mathrm{c}$  light   curve  of   \object{\wga}  is
presented in the lower panel.  The  data of Norton et al.  (\cite{Norton}) 
is marked with  open  circles. Full circles  are  from  Zharikov et al.
(\cite{Zharikov}). The AFD (all filter data) folded in the same manner is shown 
in the top panel.}
\label{fig5}
\end{figure}

  Adding  the  data kindly  provided  by  authors  of Norton  et  al. 
(\cite{Norton}) to our measurements, we were able to improve slightly the
period estimate.   The new  value for the  period of  the Intermediate
Polar \wga\, now stands at $4\fhour35\pm0\fhour01$,   similar to our
recently reported  value (Zharikov et al.  \cite{Zharikov}).  We note
that this analysis does not  change our previous estimates of the system
parameters,  but shifts  the photometric  minimum in  the  light curve
exactly to  the redefined  epoch $T_{\mathrm 0} =  2451762.9527\pm0.0001$, which
corresponds to the\ \ $\pm$ zero crossing of the \ion{H$_\beta$}{} radial
velocity  curve, i. e.   to the  moment when  the secondary  is located
between the observer and  the WD.  The final phase-folded  light curves in
the R band,  AFD, and radial velocity curves  in \ion{He}{ii} 4686 and
\ion{H$_\beta$}{} are presented in Figure \ref{fig5}. The difference of
amplitudes and phases of the \ion{H$_\beta$}{} and  \ion{He}{ii} lines were
discussed in our previous paper.

\begin{acknowledgements}  
   
This work was  supported in part by CONACYT  projects 25454-E, 36585-E
and DGAPA project IN-118999.
We are grateful to the referee A. Norton for 
the detailed comments which helped to
improve the presentation. 
      
\end{acknowledgements}

\end{document}